\documentclass[twoside,leqno]{article}

\pdfoutput=1

\usepackage[letterpaper]{geometry}

\usepackage[caption=false]{subfig}
\usepackage{ltexpprt}
\usepackage{hyperref}
\usepackage{graphicx}
\usepackage{mathtools}
\usepackage{relsize}
\usepackage{booktabs}
\usepackage{multirow}
\usepackage{makecell}
\usepackage{color}
\usepackage{xcolor}
\usepackage{amssymb}
\usepackage{amsmath}
\usepackage[sort]{cite}

\usepackage{listings}
\lstdefinestyle{customcpp}{
  breaklines=true,
  xleftmargin=\parindent,
  language=C++,
  showstringspaces=false,
  basicstyle=\small\ttfamily,
  keywordstyle=\bfseries\color{green!40!black},
  commentstyle=\itshape\color{purple!40!black},
  identifierstyle=\color{blue},
  stringstyle=\color{orange},
  tabsize=2,
  escapeinside={/@}{@/},
  keepspaces=true,
  captionpos=b,
  numbers=left,
}
\newcommand{\aeq}{\kern.35em\text{\small+}\kern-.35em=}

\newtheorem{requirement}{Requirement}

\begin{document}

\title{\Large Local Adjoints for Simultaneous Preaccumulations with Shared Inputs}
\ifpdf
\author{Johannes Bl{\"u}hdorn\thanks{Chair for Scientific Computing, University of Kaiserslautern-Landau (RPTU)}
\and Nicolas R.~Gauger\footnotemark[1]}
\else
\author{Johannes Bl{\"u}hdorn\thanks{Chair for Scientific Computing, University of Kaiserslautern-Landau (RPTU)}
\and Nicolas R.~Gauger\thanks{Chair for Scientific Computing, University of Kaiserslautern-Landau (RPTU)}}
\fi

\date{}

\newcommand{\remember}[1]{\textcolor{red}{(#1)}}

\maketitle

\begin{abstract} \small\baselineskip=9pt In shared-memory parallel automatic differentiation, inputs that are shared among simultaneous thread-local preaccumulations lead to data races if Jacobians are accumulated with a single, shared vector of adjoint variables. In this work, we discuss the benefits and tradeoffs of re-enabling such preaccumulations by a transition to suitable local adjoints. We propose different vector- and map-based approaches for storing local adjoint variables and analyze them with respect to memory consumption, memory allocation, and adjoint variable access times in the context of simultaneous preaccumulations in multiple threads. We implement the approaches in CoDiPack and benchmark them in parallel discrete adjoint computations in the multiphysics simulation suite SU2.
\end{abstract}

\section{Introduction}
\label{sec:introduction}

The multiphysics simulation suite SU2 \cite{EconomonPCTA2016} features discrete adjoints by means of operator overloading automatic differentiation (AD) \cite{AlbringSG2015}. To this end, the AD tool CoDiPack \cite{SagebaumAG2019} was applied to SU2, together with MeDiPack \cite{SagebaumEtAl2023} for the differentiation of MPI parallelism \cite{MPI2023}. SU2 implements a reverse accumulation AD workflow \cite{Christianson1994} that makes heavy use of preaccumulation to reduce the memory consumed by tapes \cite{AlbringSG2016}.\footnote{AD-specific terminology is explained in Section \ref{sec:background}.} In the course of \cite{GomesEP2021}, the MPI parallelism of SU2's primal solvers was extended by an additional layer of OpenMP parallelism \cite{OpenMP52}, allowing MPI-OpenMP hybrid parallel primal execution. By applying the OpDiLib tool for differentiation of OpenMP \cite{BluehdornSG2023}, we recently complemented this by support for hybrid parallel discrete adjoints \cite{BluehdornPAG2024}, and were able to demonstrate improved memory scaling compared to pure MPI approaches. As part of \cite{BluehdornPAG2024}, we also observed data races for simultaneous thread-local preaccumulations with shared inputs, and therefore resorted to disabling preaccumulations in the presence of sharing. This increases the memory consumption by a case-dependent offset, which counteracts otherwise improved memory scaling and motivates us to revisit and re-enable preaccumulations with sharing. In continuation of the research conducted in \cite{BluehdornPAG2024}, we propose using thread-local adjoint variables to perform simultaneous preaccumulations with shared inputs in a race-free manner. In a broader sense, this work is related to research on AD of parallel computer programs \cite{SchanenNHU2010,FoersterNU2011,Foerster2014,KalerSXLCPK2021,HueckelheimH2022,MosesEtAl2022} as well as applications of and methodology for AD of parallelized flow solvers \cite{TowaraSN2015,HueckelheimHSM2018,HueckelheimHSM2019}.

The structure of this paper is as follows. Section \ref{sec:background} provides background on AD and preaccumulation and establishes formalism and notation. In Section \ref{sec:preacc_with_shared_vars}, we identify and characterize the data race that arises in simultaneous preaccumulations with shared inputs, and suggest local adjoint variables as a promising solution. We discuss implementation approaches for local adjoint variables in Section \ref{sec:preacc_with_local_adjoints}. Section \ref{sec:implementation} introduces the demonstrator code that accompanies this paper and covers implementational aspects in CoDiPack and SU2. Section \ref{sec:benchmarking} features detailed performance studies that demonstrate the tradeoffs and benefits of re-enabled preaccumulations in SU2. We conclude our work in Section \ref{sec:conclusion}.

\section{Background}
\label{sec:background}

\begin{figure}
  \centering
  \includegraphics[width=0.48\textwidth]{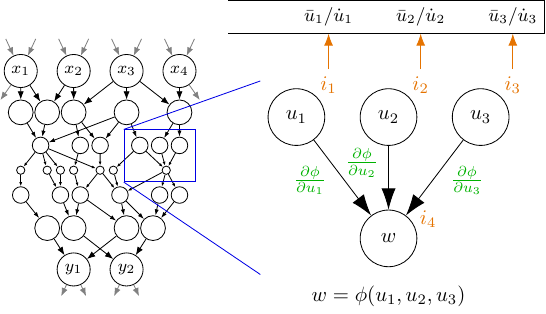}
  \caption{Computational graph in terms of statements with $n=4$ and $m=2$. Nodes are annotated with identifiers (orange) and edges are annotated with partials (green). In the context of preaccumulation, gray edges indicate connections to other parts of the graph.}
  \label{figure:computational_graph}
\end{figure}

To begin with, we briefly summarize the essentials of AD, challenges in the transition to shared-memory parallel AD, aspects of tape-based operator overloading implementations, and the idea of preaccumulation. Comprehensive introductions to AD are given, e.\,g., in \cite{GriewankW2008} or \cite{Naumann2011}. Consider a computer program $F$ that takes floating-point values $x\in\mathbb{R}^n$ as inputs and produces floating-point values $y\in\mathbb{R}^m$ as outputs. During the execution of $F$ for specific inputs $x$, each executed statement produces a new value from previous values via some composite right hand side function. The statements thus establish a dependency relation between values that can be expressed as a directed acyclic graph $G=(V,E)$ with nodes $V$ associated with the values and edges $E$ that express the direct dependencies via statements, referred to as computational graph. An example is visualized in Figure \ref{figure:computational_graph}.
To compute derivatives of $F$, AD applies the chain rule according to the structure provided by the computational graph. To this end, the edges in the computational graph are annotated with partial derivatives of the corresponding statement's right hand side function as indicated in Figure \ref{figure:computational_graph}. The forward mode of AD introduces for each value $u$ a corresponding tangent value $\dot{u}$, and the reverse mode complements each value $u$ with an adjoint value $\bar{u}$. On the level of a statement $w=\phi(u_1,\ldots,u_k)$, forward and reverse mode perform the respective updates\footnote{To account for reuse of adjoint memory locations and statements where the same variable appears on the left and right hand side, one typically creates a temporary copy of $\bar{w}$, followed by $\bar{w}=0$, and the temporary copy is used for the update \eqref{eq:reverse_statement}.}
\begin{align}
  \dot{w}&=\sum_{i=1}^k\frac{\partial\phi}{\partial u_i}(u)\dot{u}_i,\label{eq:forward_statement}\\
  \bar{u}_i&\aeq\frac{\partial\phi}{\partial u_i}(u)\bar{w}\text{~~for }i=1,\ldots,k,\label{eq:reverse_statement}
\end{align}
which, if applied throughout the computational graph, yield for the whole program respective derivatives
\begin{align}
  \dot{y}&=\frac{\partial}{\partial x}F(x)\dot{x},\label{eq:forward_ad}\\
  \bar{x}&=\frac{\partial}{\partial x}F(x)^\mathrm{T}\bar{y}\label{eq:reverse_ad}
\end{align}
with respective user-provided seeds $\dot{x}$ and $\bar{y}$. The whole Jacobian $\frac{\partial}{\partial x}F(x)$ can be assembled by multiple evaluations of \eqref{eq:forward_ad} or \eqref{eq:reverse_ad} with unit vector seeds, where the reverse mode is usually faster if $n>m$. The adjoint updates \eqref{eq:reverse_statement} depend on information that is only available after the primal computation, and compared to the primal data flow, the data flow on adjoint variables is reversed. Therefore, capturing the whole computational graph is particularly important for the reverse mode of AD, and is also referred to as recording. Tape-based operator overloading implementations of AD like CoDiPack, Adept \cite{Hogan2014}, or ADOL-C \cite{Walther2009} achieve this by replacing the floating point computation type by an AD-specific type that pairs each primal value with an identifier. Overloads of the elementary operations like $+$, $\cdot$, or $\sin$ for the AD-specific type manage (assign) the identifiers \cite{SagebaumBG2020}, capture the computational graph in terms of these identifiers, precompute the partials $\frac{\partial\phi}{\partial u_i}(u)$ (specifically in Jacobian taping), and store everything in a stack-like structure called tape (the data layout is discussed, e.\,g., in \cite{Hogan2014}). During derivative computations (subsequently also referred to as evaluation), the identifiers serve as virtual addresses to access vectors of tangent or adjoint values \cite{NaumannR2006}, as also visualized in Figure \ref{figure:computational_graph}. We rely on a parallel reuse strategy to manage identifiers.

In shared-memory parallel reverse AD, each thread records its computations on its own thread-local tape, and a differentiation logic for OpenMP such as provided by OpDiLib ensures the correct handling of OpenMP parallelism and synchronization during the recording and provides correspondingly parallelized and synchronized reverse evaluations with a single, shared vector of adjoint variables. With this shared vector, simultaneously executed primal statements with a shared variable $u_i$ on their right hand sides produce conflicting simultaneous updates \eqref{eq:reverse_statement} on the adjoint variable $\bar{u}_i$. The known approaches to resolve these inherent data races include atomic operations \cite{FoersterNU2011,Foerster2014,HueckelheimHSM2019} and reductions \cite{GieringKTEGW2006,Kowarz2008,BischofGKW2008,KalerSXLCPK2021,HueckelheimH2022}.

Tapes can quickly exceed available memory capacities. Preaccumulation aims at reducing this memory consumption and targets subgraphs of the computational graph that satisfy Requirement \ref{req:one}.
\begin{requirement}
\label{req:one}
Preaccumulated subgraphs connect to the remaining computational graph only via incoming/outgoing edges to/from the subgraph's inputs and outgoing edges from the subgraph's outputs.
\end{requirement}
This is visualized in Figure \ref{figure:computational_graph}. As explained e.\,g.~in \cite{GriewankW2008}, if such a subgraph corresponding to some program part $P$ has relatively few inputs and outputs compared to the number of intermediate nodes and edges, the precomputed Jacobian of $P$ together with information on the inputs and outputs of $P$ consumes less memory than the usual tape recording of $P$. In terms of the thus modified recording of $F$, preaccumulation trades increased recording time (Jacobian computation of $P$) for reduced evaluation time (smaller tape for $F$). This is particularly appealing in a reverse accumulation setting \cite{Christianson1994}, where a tape is recorded once but evaluated multiple times.
Successful applications of preaccumulation include precomputation of Jacobians on the statement level (the partials $\frac{\partial\phi}{\partial u_i}(u)$ in \eqref{eq:forward_statement} and \eqref{eq:reverse_statement}, noting that $\phi$ is usually composed of multiple elementary operations) \cite{BischofCCG1992,PhippsP2012,Hogan2014}, Jacobian computations of basic blocks in source transformation AD \cite{Utke2006}, and Jacobian computations of larger code sections that have been marked together with their inputs and outputs as targets for preaccumulation according to developer knowledge \cite{LotzNU2012,AlbringSG2016}. The latter is the type of preaccumulation that we consider in this work.

Finding a preaccumulation procedure for a given computational graph that produces the Jacobian with a minimal number of arithmetic operations is known as the Optimal Jacobian Accumulation problem, which is known to be NP-complete \cite{Naumann2008}. Elimination techniques on the computational graph (vertex and edge elimination) or its dual graph (face elimination) may serve as tools towards improved solutions \cite{Naumann2004} and optimal strategies are known for special classes of graphs \cite{NaumannH2008}. In this work, we use standard forward and reverse AD according to \eqref{eq:forward_statement} and \eqref{eq:reverse_statement} to perform preaccumulations, which can be interpreted as certain sequences of back- and front-elimination of edges, respectively \cite{GriewankW2008}. CoDiPack, for instance, first records $P$ in continuation of the already recorded program parts prior to $P$. The recording of $P$ is subsequently evaluated in the spirit of \eqref{eq:forward_ad} or \eqref{eq:reverse_ad} (depending on the number of inputs and outputs) with unit vector seeds, possibly multiple times, until the full Jacobian of $P$ is assembled. Finally, the recording of $P$ is discarded and replaced by the Jacobian together with information on the inputs and outputs of $P$. Naturally, the AD tool reuses the facilities that perform the overall tape recording and evaluation also for the assembly of Jacobians in local preaccumulations. In particular, tape evaluations for local preaccumulations reuse the global shared vector of adjoint variables and, depending on whether preaccumulation is performed with the forward or reverse mode of AD, use it to store their tangent or adjoint values. This introduces additional challenges in a shared-memory setting.

\section{Preaccumulations with Shared Inputs}
\label{sec:preacc_with_shared_vars}

By availability of a differentiation logic for OpenMP and appropriate treatment of inherent data races in the reverse pass, the implementation of preaccumulation described in Section \ref{sec:background} extends readily to a \emph{single} preaccumulation workflow where the preaccumulated code contains OpenMP parallel parts. This study focusses on simultaneous execution of \emph{multiple} preaccumulation workflows, which arises for example if preaccumulation is applied inside items of an OpenMP worksharing loop. For the problem identification, let $G_1$ and $G_2$ denote two computational subgraphs corresponding to program parts $P_1$ and $P_2$ that should be preaccumulated simultaneously (all of the following extends to arbitrary numbers of simultaneously preaccumulated subgraphs). Requirement \ref{req:one} implies that $G_1$ and $G_2$ neither share intermediate nor output nodes, but may have input nodes in common. Likewise, simultaneously executed non-preaccumulated parts of the computational graph may read the inputs of $G_1$ and $G_2$, but neither read nor write intermediate or output nodes of $G_1$ and $G_2$ (assuming that the primal program is free of data races).\footnote{A data access pattern that only allows shared inputs among threads is also a requirement in the differentiation approach for parallel regions and loops by Kowarz \cite{Kowarz2008} and its implementation in ADOL-C \cite{KowarzW2008}.} Since shared read access does not require synchronization, a reasonable extension of Requirement \ref{req:one} in the shared-memory parallel case is found by Requirement \ref{req:two}.
\begin{requirement}
\label{req:two}
There is no synchronization dependency between a preaccumulated subgraph and the remaining computational graph.\footnote{Synchronization might make it challenging (maybe impossible) to perform preaccumulations in a correct manner and to retain appropriate synchronization when embedding the Jacobians into the surrounding computational graph; to give an artificial example, consider a barrier directive that some threads of a team encounter inside preaccumulated code, while others encounter it outside of preaccumulated code.}
\end{requirement}
In particular, there is no synchronization between $P_1$ and $P_2$, and no synchronization between either $P_1$ or $P_2$ and the remaining program parts. However, synchronization \emph{inside} $P_1$ or $P_2$, e.\,g.~among the threads of a a nested parallel region, is allowed. We remark that the OpenMP-parallel version of SU2 satisfied Requirement \ref{req:two} without explicit revision of synchronization constructs \cite{BluehdornPAG2024}.

\begin{figure*}[t]
  \begin{center}
    \subfloat[]{\label{figure:independent_preacc}\includegraphics[scale=0.8]{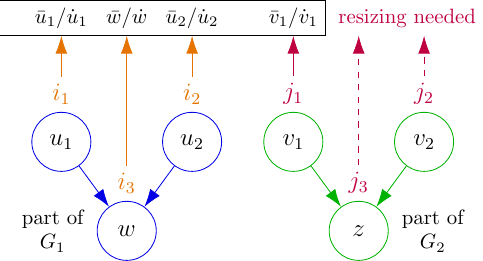}}
    \hfill
    \subfloat[]{\label{figure:shared_input_preacc}\includegraphics[scale=0.8]{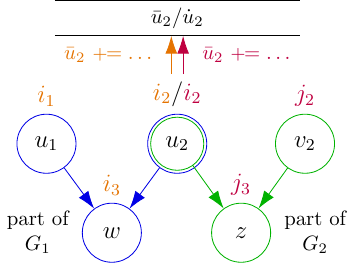}}
    \hfill
    \subfloat[]{\label{figure:local_adjoints_preacc}\includegraphics[scale=0.8]{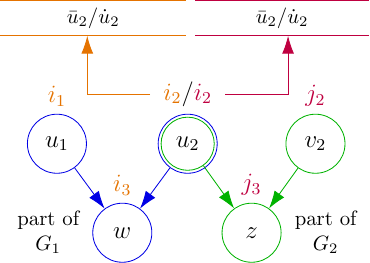}}
  \end{center}
  \caption{Preaccumulation involves resizing of the shared vector of adjoint/tangent variables, which requires mutual exclusion with tape evaluations in concurrent preaccumulations (a). Simultaneous preaccumulations with shared inputs lead to data races on the shared vector of adjoints/tangents (b). We propose using thread-local memory to mitigate these data races (c).}
\end{figure*}

With Requirement \ref{req:two}, the implementation of preaccumulation described in Section \ref{sec:background} also extends to simultaneous preaccumulations without shared inputs, with the caveat that the shared vector of adjoint variables needs to be managed carefully. This is visualized for $G_1$ and $G_2$ in Figure \ref{figure:independent_preacc}. As the recording proceeds, new, larger identifiers are created, which do not yet correspond to a memory location in the shared vector of adjoint variables. Prior to evaluations in the context of preaccumulation, the vector of adjoint variables has to be resized accordingly. To avoid data races, a shared mutex ensures mutual exclusion between evaluations and resizing. Note that the corresponding locking can limit the scalability of simultaneous preaccumulations.

The situation that $G_1$ and $G_2$ share an input node $u_2$ is depicted in Figure \ref{figure:shared_input_preacc}. A computation of the Jacobian with the reverse mode of AD accumulates a Jacobian entry in the adjoint variable $\bar{u}_2$; a computation of the Jacobian with the forward mode of AD attempts to set a seed $\dot{u}_2=1$; both cases involve respective resets of $\bar{u}_2$ and $\dot{u}_2$ to $0$. Either combination of modes for simultaneous preaccumulation of $G_1$ and $G_2$ (both forward; both reverse; one forward, one reverse) leads to data races on the shared memory location $\bar{u}_2/\dot{u}_2$. Like the inherent data races in shared-memory parallel reverse AD, these races originate from non-exclusive read access. Unlike the inherent data races, however, these data races are not resolved by mechanisms that perform modifications of $\bar{u}_2/\dot{u}_2$ in a race-free manner. Suppose for example that both $G_1$ and $G_2$ are preaccumulated with the reverse mode of AD and consider the updates \eqref{eq:reverse_statement} to $\bar{u}_2$; this is also visualized in Figure \ref{figure:shared_input_preacc}. Even if these updates were performed atomically, $\bar{u}_2$ would in the end still hold the sum of two unrelated entries of the Jacobians of $G_1$ and $G_2$.\footnote{As the preaccumulation workflows of $G_1$ and $G_2$ are not synchronized with respect to each other and might involve multiple tape evaluations and adjoint variable resets, $\bar{u}_2$ actually changes in an unpredictable manner.} Instead, our goal is to avoid the mixture of contributions from different preaccumulations entirely.

In addition to the usual benefits in terms of memory consumption and evaluation performance that we expect from preaccumulation, finding ways to enable preaccumulations with shared inputs seems particularly attractive: without preaccumulation, the non-exclusive read access results in inherent data races (see Section \ref{sec:background}), so that the non-preaccumulated recordings would have to be evaluated with mitigations for data races in the reverse pass, like atomic updates. With preaccumulation, on the other hand, these atomic operations are eliminated from the tape, so that the expected performance benefit for the tape evaluation is larger. A developer may convert preaccumulations with shared inputs into preaccumulations with mutually independent inputs manually by privatizing shared variables prior to registration as preaccumulation inputs. With identifier management strategies that assign different identifiers to copies, this is achieved by creating thread-local copies of the respective input variables. In addition, all references to respective inputs throughout the preaccumulated code have to be replaced by references to the corresponding thread-local copies. This modification is not always feasible or convenient in practice, and not automatic either. A general solution is found in the transition to thread-local memory for adjoint/tangent variables associated with shared inputs. As visualized in Figure \ref{figure:local_adjoints_preacc}, the data race on $\bar{u}_2/\dot{u}_2$ from \ref{figure:shared_input_preacc} is addressed by providing thread-local storage for $\bar{u}_2/\dot{u}_2$. This is similar to the reduction approach for addressing the inherent data races in shared-memory parallel reverse AD in the sense that we also use thread-local memory, but different in the sense that we do not perform any reductions. For a unified and consistent implementation, it seems natural to extend this to all adjoint/tangent memory locations of a preaccumulation, not only those associated with inputs, and perform preaccumulation entirely on collections of thread-local adjoint/tangent variables.\footnote{For brevity, we just speak of local adjoints in the following.} With a view on the shared-memory parallel setting, local adjoints also improve memory locality.

\section{Preaccumulation with Local Adjoints}
\label{sec:preacc_with_local_adjoints}

Let $G_t=(V_t,E_t)$, $t=1,\ldots,T$ be a family of computational subgraphs for simultaneous preaccumulation, each to be performed by one of $T$ threads. The $G_t$ may share common input nodes. Let $\operatorname{idf}_{G_t}\colon V_t\to\mathbb{N}$ denote the identifiers attached to the nodes of $G_t$. To preaccumulate each $G_t$ with local adjoints, we need a thread-local data structure for local adjoints that can be accessed with the respective identifiers $\operatorname{idf}_{G_t}(V_t)$. The choice of data structure impacts the preaccumulation performance in multiple ways.
\begin{enumerate}
  \item The time spent on \emph{allocating and zero-initializing local adjoints} adds to the preaccumulation time. Allocations/initializations in different threads can be performed in parallel.\footnote{\label{footnote:memory_bandwidth}In practice, scaling can only be expected until $T$ becomes large enough to saturate the memory bandwidth.}
  \item The data structure for local adjoints determines \emph{adjoint access times}. Local adjoint accesses in different threads can be performed in parallel.\textsuperscript{\ref{footnote:memory_bandwidth}}
  \item The \emph{cumulative memory spent on local adjoints} by simultaneous preaccumulations adds to the overall memory consumption.
\end{enumerate}
In the following, we propose different storage strategies for local adjoints and assess them with respect to these criteria. For a simplified analysis, we assume that the graphs $G_t$ are isomorphic, that is, they are identically structured but work on different data with different identifiers (with the exception of shared inputs), and we assume that the structure of the $G_t$ does not depend on $T$. This situation arises for example if the $G_t$ stem from items of the same worksharing loop, where $T$ only determines the number of items that are processed in parallel. We furthermore assume for the analysis that no identifiers are reused, i.\,e.~$|V_t|=|\operatorname{idf}_{G_t}(V_t)|$; our results still serve as upper bounds otherwise. With these assumptions, we are able to argue about how the adjoints-related memory consumption depends on $T$.

\begin{figure*}[t]
  \begin{center}
    \subfloat[]{\label{figure:local_vector}\includegraphics[scale=0.8]{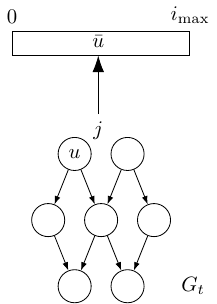}}
    \hfill
    \subfloat[]{\label{figure:local_vector_offset}\includegraphics[scale=0.8]{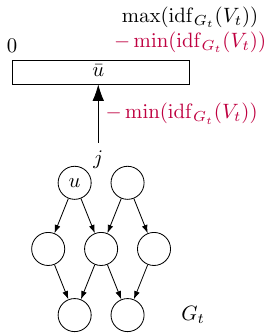}}
    \hfill
    \subfloat[]{\label{figure:local_map}\includegraphics[scale=0.8]{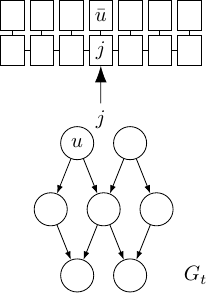}}
    \hfill
    \subfloat[]{\label{figure:identifier_map}\includegraphics[scale=0.8]{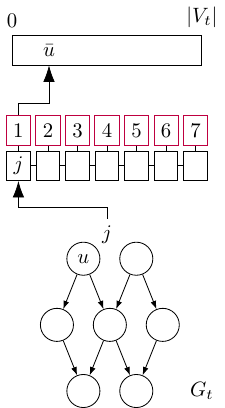}}
    \hfill
    \subfloat[]{\label{figure:tape_editing}\includegraphics[scale=0.8]{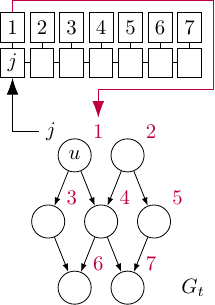}}
  \end{center}
  \caption{Implementation strategies for local adjoints. Thread-local duplicates of the global adjoint vector (a), addressing into these vectors with an offset (b), map-based local adjoints (c), remapped identifiers with small, dense thread-local vectors (d), tape editing to accelerate multiple such evaluations (e).}
\end{figure*}

\subsection{Vector-Based Approaches}

As a natural extension of preaccumulation on the global vector of adjoint variables, we begin by assessing vector-based storage strategies for local adjoints, e.\,g.~\lstinline{std::vector<double>} or an array \lstinline{double[]} \cite{IsoCpp2011}. The reduction approach for resolving inherent data races of shared-memory parallel reverse AD as implemented, e.\,g., in \cite{Kowarz2008,KowarzW2008} duplicates the entire recording environment per thread, including the global vector of adjoint variables. Inspired by this, we may provide thread-local vectors of adjoint variables that are used by preaccumulations instead of the global vector of adjoint variables. This is visualized in Figure \ref{figure:local_vector}. This approach retains the $\mathcal{O}(1)$ times for individual adjoint accesses. In order to be accessible with any identifier up to and including $\max(\operatorname{idf}_{G_t}(V_t))$, a local adjoint vector for the preaccumulation of $G_t$ requires a size of $\max(\operatorname{idf}_{G_t}(V_t))+1$. To avoid the allocation and initialization of an entire local adjoint vector for each individual preaccumulation, we use persistent thread-local adjoint vectors that are resized as needed. Let $i_{\mathrm{max}}$ denote the maximum identifier attached to any node throughout the $G_t$ as well as any node in program parts recorded prior to the $G_t$. $i_{\mathrm{max}}$ is usually known to the identifier management, and without computing $\max(\operatorname{idf}_{G_t}(V_t))$, local adjoint vectors may directly be resized to $i_{\mathrm{max}}+1$. This leads to a cumulative memory consumption of $\mathcal{O}(i_{\mathrm{max}}\cdot T)$, and the allocation/initialization effort in each thread ranges from zero (no resizing needed) to $\mathcal{O}(i_{\mathrm{max}})$ (reallocation of the entire vector). An upper bound for the memory consumption due to local adjoints is given by $\mathcal{O}(i^{\mathrm{global}}_{\mathrm{max}}\cdot T)$, where $i^{\mathrm{global}}_{\mathrm{max}}$ is the maximum identifier used throughout the entire recording. As preaccumulations may occur arbitrarily late in the AD workflow, this memory consumption may add to the memory consumption of the entire tape.

To improve the memory consumption, we may determine $\min(\operatorname{idf}_{G_t}(V_t))$ and $\max(\operatorname{idf}_{G_t}(V_t))$ in $\mathcal{O}(|V_t|)$ time by visiting every node in $G_t$ once. Instead of using the identifiers in $G_t$ directly to address into the local adjoint vector, we first subtract $\min(\operatorname{idf}_{G_t}(V_t))$, so that a local adjoint vector of size $\max(\operatorname{idf}_{G_t}(V_t)) - \min(\operatorname{idf}_{G_t}(V_t)) + 1$ is sufficient, with a corresponding improvement in the memory consumption and allocation/initialization workload. This is visualized in Figure \ref{figure:local_vector_offset}. The improvement depends on the distance of $\min(\operatorname{idf}_{G_t}(V_t))$ and $\max(\operatorname{idf}_{G_t}(V_t))$; in the worst case $\min(\operatorname{idf}_{G_t}(V_t))=1$\footnote{The identifier $0$ usually serves as an identifier for passive data and does not appear in a recorded graph.} and $\max(\operatorname{idf}_{G_t}(V_t))=i_{\mathrm{max}}$, there is essentially no improvement. This may realistically happen in practice if variables with small identifiers, e.\,g.~inputs of the global AD workflow, are registered as preaccumulation inputs, noting that the recording for the preaccumulation itself usually creates new identifiers, that are usually among the largest identifiers created so far.

The major downside of these vector-based approaches is that large, dense data structures are duplicated per thread but only ever hold sparse data, which counteracts a primary objective of preaccumulation: reducing the memory consumption. As the additional memory consumption due to local adjoints increases linearly with $T$, this also counteracts the improvements in terms of memory scaling due to the transition from pure MPI to a combination of MPI and OpenMP that we demonstrated in \cite{BluehdornPAG2024}.

\subsection{Map-Based Approaches}
\label{section:map_based_approaches}

We continue with assessing associative containers that better reflect the sparsity of local adjoints. Specifically, we consider \lstinline{std::map<int, double>} and \lstinline{std::unordered_map<int, double>} \cite{IsoCpp2011}. Both can provide $|V_t|$ memory locations for local adjoints accessible with the identifiers $\operatorname{idf}_{G_t}(V_t)$ with a memory consumption of $\mathcal{O}(|V_t|)$ independently of the specific identifiers used in $G_t$, resulting in a temporary additional memory consumption of $\mathcal{O}(|V_t|\cdot T)$ during preaccumulation of the family $G_t$. This is visualized in Figure \ref{figure:local_map}. For both \lstinline{std::map<int, double>} and \lstinline{std::unordered_map<int, double>}, \lstinline{operator[]} allocates and default-initializes adjoint memory accessible with the provided identifier if needed, and returns a reference to this adjoint memory. This means that the first tape evaluation in the course of preaccumulating $G_t$ can start with an empty map and populate it on the fly, and there is no need to perform preprocessing on the identifiers $\operatorname{idf}_{G_t}(V_t)$. For \lstinline{std::map<int, double>}, the time complexity for an individual insertion or element access is logarithmic in the current size of the container, i.\,e.~at most $\mathcal{O}(\log|V_t|)$. For \lstinline{std::unordered_map<int, double>}, the average time complexity for an individual insertion or element access is constant, i.\,e.~$\mathcal{O}(1)$, but the worst case time complexity is linear in the current size of the container, i.\,e.~at most $\mathcal{O}(|V_t|)$.

Map-based approaches appeal in particular due to low memory consumption, but achieve it at the price of increased adjoint access times.

\subsection{Identifier Preprocessing}

Instead of providing local mapped adjoints for the preaccumulation of $G_t$, we may also use analogous containers \lstinline{std::map<int, int>} or \lstinline{std::unordered_map<int, int>} to remap the identifiers $\operatorname{idf}_{G_t}(V_t)$ to the contiguous range $\{1,\ldots,|V_t|\}$, and use the thus remapped identifiers to access a local adjoint vector of size $|V_t|+1$. This is visualized in Figure \ref{figure:identifier_map}. The identifier map can be populated and accessed by means of the \lstinline{insert} method, to which we pass the original identifier paired with the next identifier in the contiguous range. If needed, it inserts the original identifier into the map and associates it with this next identifier in the contiguous range. \lstinline{insert} returns an iterator (reference) to the newly inserted or already existing element together with indication whether insertion was performed, in which case we increment the next identifier in the contiguous range. For both types of map, the \lstinline{insert} method has the same time complexity as the respective \lstinline{operator[]}. If identifiers are remapped like this during tape evaluations, we achieve the same map-related time and memory complexity as with mapped adjoints with the analogous map type, yet with higher total memory consumption as we spend memory on both an identifier map and a local adjoint vector, and there is an additional indirection in the adjoint access. We therefore expect thus remapped identifiers to perform worse than mapped adjoints. Unlike adjoints, however, identifiers can be remapped up-front. Prior to tape evaluations for $G_t$, we iterate over all identifiers in the tape of $G_t$. We assemble the identifier map in the same fashion as explained above, and use it at the same time to edit the tape and replace the identifiers by identifiers from the contiguous range $\{1,\ldots,|V_t|\}$. This is visualized in Figure \ref{figure:tape_editing}. Each tape evaluation can then be performed with $\mathcal{O}(1)$ adjoint access times on a local adjoint vector of size $|V_t|+1$. If the association between original input and output identifiers and Jacobian entries is made by position, the identifier map can be discarded prior to the allocation of the local adjoint vector.

This approach combines some of the advantages of the vector-based and map-based strategies and helps to reduce the runtime overhead of map-based approaches. The time complexity of identifier preprocessing for $G_t$ with one of the two map types is identical to the joint time complexity of all adjoint accesses in the first tape evaluation when preaccumulating $G_t$ with local mapped adjoints and the analogous map type. This means that we can expect identifier preprocessing to perform better than mapped adjoints for preaccumulations that involve multiple tape evaluations, i.\,e., have at least two inputs and two outputs.

\section{Implementation}
\label{sec:implementation}

We provide a minimal demonstrator code to illustrate the ideas of this paper.\footnote{\url{https://github.com/SciCompKL/local_adjoints_demonstrator}} The code features exemplary implementations of the various approaches for local adjoints discussed in Section \ref{sec:preacc_with_local_adjoints} and applies them to synthetic preaccumulation workloads resembling tapes for single-input single-output chains of unary operations. The code can be used for quick exploration and benchmarking purposes. For further details, please refer to the documentation in the repository and the comments in the code.

To implement preaccumulation with local adjoints in CoDiPack, we extend available functionality for tape evaluations with custom adjoints vectors, so that the respective methods accept generalized collections of custom adjoint variables instead, and make corresponding changes to CoDiPack's internal evaluation routines. To give a qualitative example, methods like
\begin{lstlisting}[numbers=none]
template<typename Gradient>
void Tape::evaluate(Gradient* adjoints);
\end{lstlisting}
become
\begin{lstlisting}[numbers=none]
template<typename Adjoints>
void Tape::evaluate(Adjoints&& adjoints);
\end{lstlisting}
with the implicit requirement that \lstinline{Adjoints::operator[]} is implemented. At the time of writing, this functionality is already available in CoDiPack's \texttt{develop} branch.\footnote{\url{https://github.com/SciCompKL/CoDiPack/tree/develop}} Based on this, we implement preaccumulation with local adjoints in CoDiPack's \lstinline{PreaccumulationHelper}. We provide a dedicated \lstinline{PreaccumulationHelper::finish*()} method for each of the implementation strategies for local adjoints. These serve as alternatives to the classical \lstinline{PreaccumulationHelper::finish()} method that uses the shared global vector of adjoint variables. At the time of writing, the strategies discussed in this paper are implemented in a dedicated branch\footnote{\url{https://github.com/jblueh/CoDiPack/tree/local_adjoints}} and are partly available in CoDiPack's \texttt{develop} branch.

In the course of implementing hybrid parallel discrete adjoints in SU2, we already analyzed and annotated SU2 with respect to shared reading of variables, for the purpose of optimizing for parts of the code that can be evaluated without atomic adjoints and to disable preaccumulation in the presence of sharing \cite{BluehdornPAG2024}. Instead of disabling preaccumulations, we now perform them on local adjoints instead. At the time of writing, a version of SU2 that can make use of local adjoints is available in a dedicated branch.\footnote{\label{footnote:su2_branch}\url{https://github.com/jblueh/SU2/tree/local_adjoints}} This branch also includes build instructions and the test cases that we use in the following benchmarks.\footnote{\label{footnote:su2_benchmark}\url{https://github.com/jblueh/SU2/tree/local_adjoints/benchmark}} It is based on the branch that accompanies our previous publication \cite{BluehdornPAG2024} and should be used together with the aforementioned CoDiPack branch.

\section{Benchmarking}
\label{sec:benchmarking}

\begin{figure*}[th!]
\includegraphics[width=1\textwidth]{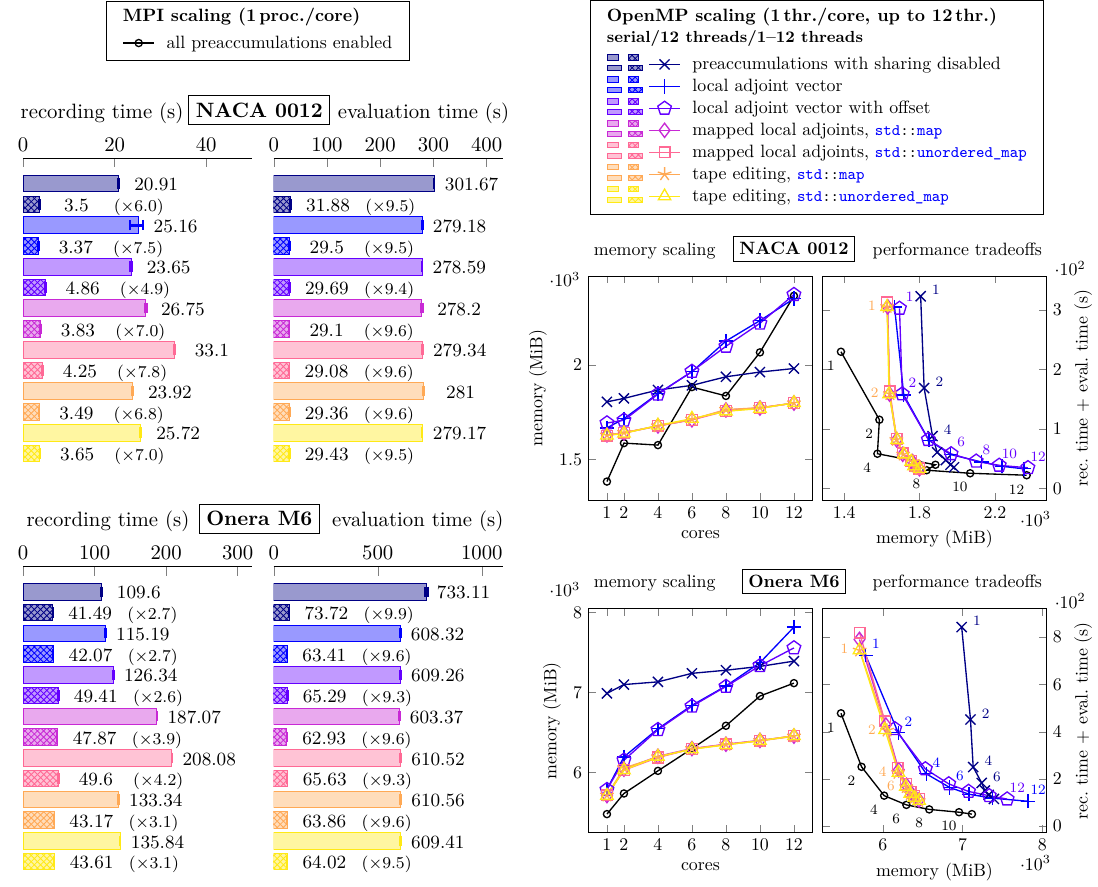}
\caption{Single-socket NACA 0012 and Onera M6 tests with different strategies for preaccumulation. On the left, we assess recording and evaluation times for serial and 12-fold OpenMP parallel execution, including speedups relative to serial execution and error bars to indicate variation across runs. On the right, we assess the memory consumption with varying degrees of parallelism as well as joint memory and runtime performance characteristics, and also include measurements of MPI-parallel runs of the classical, MPI-only build of SU2.}
\label{fig:single_socket}
\end{figure*}

We evaluate the performance of preaccumulations with local mapped adjoints in hybrid parallel discrete adjoint computations in SU2, in continuation of the performance studies conducted in \cite{BluehdornPAG2024}, using the same test cases from external aerodynamics. To summarize, we work with a NACA 0012 geometry (75140 quadrilaterals) and an Onera M6 geometry (258969 tetrahedrons) for single-socket performance tests, and a mesh of the NASA Common Research Model (HL-CRM) for performance tests with multiple nodes (approx.~18 million cells and 8.3 million nodes, mixed element types). For each test case, we start with a converged primal flow solution and execute a fixed number of discrete adjoint iterations according to SU2's reverse accumulation AD workflow (300, 300, and 1000 iterations, respectively). To ensure that workloads across different parallel setups are comparable, special measures are in place, such as reduced tolerances to avoid early stopping and disabling of special behaviour for OpenMP parallel runs with a single thread. For further details, please refer to \cite{BluehdornPAG2024} or the accompanying SU2 branch.\textsuperscript{\ref{footnote:su2_branch},\ref{footnote:su2_benchmark}}

We work with dual-socket Skylake nodes of the Elwetritsch cluster at the RPTU, that feature Intel Xeon Gold 6126 processors, each with a single NUMA domain consisting of twelve cores. We fix the clock frequency to 2.6\,GHz. Within each job allocation, we perform a discarded warm-up run, followed by five benchmark runs whose timings are averaged and displayed together with their variation in our performance plots. For single-socket tests, we use one MPI process, bound to a single socket, with one to twelve OpenMP threads. For multi-node tests, we vary the number of nodes. We use two MPI processes per node, each bound to one of the sockets, and twelve OpenMP threads per MPI process. We also include performance results obtained with a classical, MPI-only build of SU2, in which we used one MPI process per core, bound to the core's socket. We build SU2 with GCC 11.3.

We compare all the different strategies for preaccumulation with shared inputs that we discussed in Section \ref{sec:preacc_with_local_adjoints}: the baseline configuration where preaccumulations with shared inputs are disabled; performing preaccumulations with shared inputs on a thread-local copy of the global adjoint vector; using an offset when addressing into this vector; mapped local adjoints with \lstinline{std::map} and \lstinline{std::unordered_map}; remapping identifiers and editing the tape with \lstinline{std::map} and \lstinline{std::unordered_map}; performing all preaccumulations on global adjoints in a classical, MPI-only build of SU2 that is not subject to the sharing issue.

The left part of Figure \ref{fig:single_socket} displays the influence of preaccumulations with local adjoints on the recording and evaluation performance of the single-socket test cases. The trends are similar for both test cases. All variants of local adjoints succeed likewise at reducing the evaluation times, but result in increased recording times, in particular at serial execution. This is in parts due to the preaccumulations with sharing that were previously disabled, but also due to the specific strategy for local adjoints. Compared to vector-based approaches, map-based approaches are notably slower. Given the analysis in Section \ref{sec:preacc_with_local_adjoints}, we attribute this to the adjoint access times. \lstinline{std::map} performs better than \lstinline{std::unordered_map} in the two test cases. Tape editing in turn reduces the overhead notably. Comparing map-based approaches with the baseline configuration, we also see that the map-related workloads scale quite well. Even though the recording times generally increase with local adjoints, note that the reduced evaluations times lead to a net runtime improvement in all serial and almost all displayed parallel executions of the two test cases. In general, this can be expected with a sufficient number of discrete adjoint iterations, i.\,e., sufficiently many global tape evaluations per global tape recording.

\begin{figure*}[th!]
\includegraphics[width=1\textwidth]{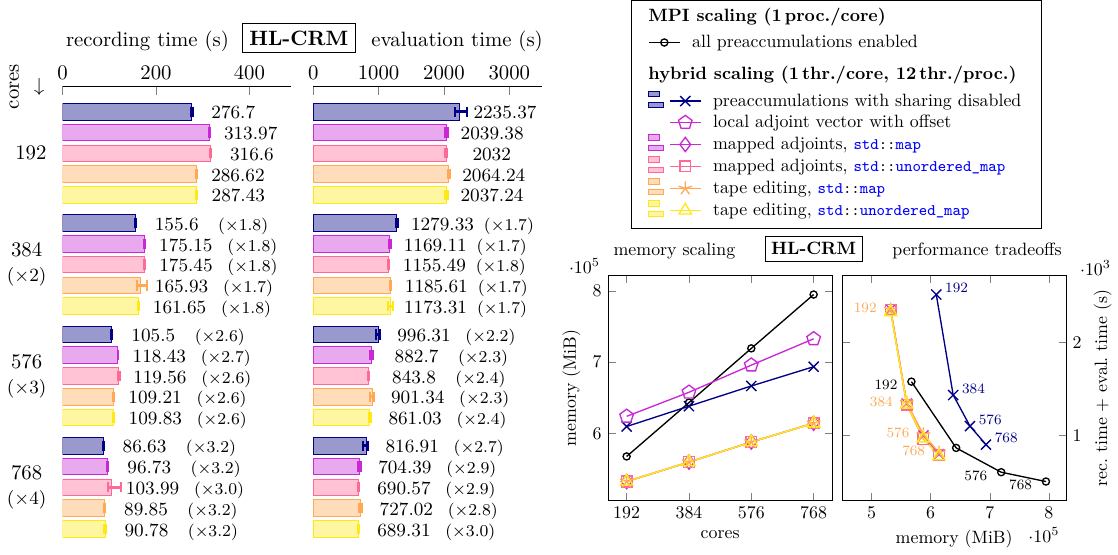}
\caption{Multi-node HL-CRM tests with different strategies for preaccumulation and various degrees of parallelism. On the left, we assess recording and evaluation times. Speedups are relative to the performance with 192 cores (eight nodes), error bars indicate variation across runs. On the right, we assess the memory consumption as well as joint memory and runtime performance characteristics, including measurements of MPI-parallel runs of the classical, MPI-only version of SU2.}
\label{fig:multi_node}
\end{figure*}

In the right part of Figure \ref{fig:single_socket}, the plots on the left show the memory consumption of the single-socket test cases with varying degrees of OpenMP parallelism. Compared to the baseline configuration, all approaches with local adjoint succeed at reducing the offset in the memory consumption at serial execution, but only map-based approaches preserve the favourable rate at which memory usage grows with an increasing number of OpenMP threads. All map-based approaches result in similar memory consumption. This is in line with the analysis in Section \ref{sec:preacc_with_local_adjoints}. The measurements also show that addressing into thread-local adjoint vectors with an offset does not necessarily result in improvements of the memory usage in practice. Compared to the the classical MPI-only approach, twelve-fold OpenMP parallelism was not sufficient for the baseline configuration to achieve lower memory consumption in the Onera M6 test case, whereas re-enabling preaccumulations with map-based approaches achieves parity already at six-fold parallelism. The rightmost plots in Figure \ref{fig:single_socket} display the joint performance characteristics of all configurations in terms of memory consumption and combined recording and evaluation time with varying degrees of parallelism. Anything further to left consumes less memory and anything further to the bottom requires less AD-specific runtime. The plots clearly illustrate how re-enabling of preaccumulations with map-based approaches helps to close the performance gap between the baseline configuration and the classical MPI-based approach. Specifically, eight-fold MPI parallelism no longer dominates larger degrees of OpenMP parallelism in the NACA 0012 case, and while all degrees of OpenMP parallelism were previously dominated by their MPI-parallel counterparts in the Onera M6 test case, larger degrees of parallelism now allow trading memory for runtime (notwithstanding, six-fold MPI parallelism still dominates higher degrees of OpenMP parallelism in the Onera M6 case). Vector-based approaches for local adjoints, on the other hand, do not lead to a clear overall improvement, especially for higher degrees of parallelism.

The corresponding performance results for the HL-CRM test case on multiple nodes are displayed in Figure \ref{fig:multi_node} and confirm the trends and benefits of re-enabled preaccumulations with map-based approaches observed in the single-socket tests, in particular the slightly reduced recording performance at twelve-fold OpenMP parallelism per MPI process, the improvements due to tape editing in this regard, the notably improved evaluation performance, and the notably reduced memory consumption. The difference between \lstinline{std::map} and \lstinline{std::unordered_map} is not as large in this test case. Configurations with re-enabled preaccumulations with map-based approaches consume less memory than the classical MPI-only approach already at 192-fold parallelism (eight nodes). The memory plot also includes results with one of the vector-based approaches, and clearly, it does not improve the memory consumption compared to the baseline configuration, which is why we otherwise do not consider it in greater detail. In terms of joint memory and runtime characteristics, the improvements due to re-enabled preaccumulations with map-based approaches result in Pareto-optimal performance in the sense that there is no classical MPI-only configuration that performs better both in terms of memory and runtime.

\section{Conclusion}
\label{sec:conclusion}

In previous work on shared-memory parallel AD, we identified data races in simultaneous preaccumulations with shared inputs, and resorted to disabling them, at the price of an offset in the memory consumption. In this work, we elaborated on the nature of these data races and proposed to re-enable preaccumulations with shared inputs by performing them on thread-local adjoint variables. We analysed various vector-based and map-based storage strategies for local adjoints, and found that map-based approaches align well with our objective of reducing the memory consumption while preserving improvements in the memory scaling due to the transition from pure MPI to a combination of MPI and OpenMP. We furthermore proposed to reduce the runtime overhead of map-based approaches by moving from mapped adjoints to a combination of mapped identifiers and tape editing. We implemented the various strategies for preaccumulations with local adjoints both in a minimal demonstrator code and in CoDiPack. Using the latter, we explored the tradeoffs and benefits of the various approaches in hybrid parallel discrete adjoint computations in SU2. We saw that preaccumulation-related overheads in the recording phase can be compensated for by a sufficient number of faster tape evaluations in a reverse accumulation AD workflow, especially in the OpenMP-parallel setting since the workloads for preaccumulations with map-based local adjoints scale very well. As we were able to demonstrate net improvements both in terms of memory and runtime with map-based approaches compared to the conservative approach of disabling preaccumulations with shared inputs, this work marks an additional step towards improved hybrid parallel AD performance, and specifically in our SU2 test cases, helps to further close the performance gap between MPI-parallel and hybrid parallel discrete adjoint computations.

\section*{Acknowledgements}

We would like to thank Uwe Naumann and the anonymous reviewers for their valuable suggestions and remarks. We gratefully acknowledge funding from the German National High Performance Computing (NHR) association for the Center NHR South-West.

\bibliographystyle{siam}
\bibliography{literature.bib}

\end{document}